\documentclass[prd,preprintnumbers,superscriptaddress,nofootinbib,twocolumn]{revtex4-2}
\usepackage{epsfig}
\usepackage{amsmath}
\usepackage{amsfonts}
\usepackage{float}
\usepackage{amssymb}
\usepackage{slashed}
\usepackage{pbox}
\usepackage{subfigure}
\usepackage[colorlinks,citecolor=black, linkcolor = black, urlcolor = black]{hyperref}
\usepackage{tabularx}
\usepackage{titlesec}
\usepackage{scrextend}
\usepackage[version=3]{mhchem}
\usepackage{tabularray}
\usepackage[normalem]{ulem}
\usepackage[bottom]{footmisc}
\usepackage{mathtools}

\def\lsim{\lesssim}

\def\be{\begin{eqnarray}}
\def\ee{\end{eqnarray}}

\def\nn{\nonumber}

\def\D{\text{D}}

\def\bal#1\eal{\begin{align}#1\end{align}}
\def\beq{\begin{equation}}
\def\eeq{\end{equation}}

\def\GeV{{\rm GeV}}
\def\MeV{{\rm MeV}}

\def\eV{{\rm eV}}

\def\H{\text{H}}

\def\He{\text{He}}

\def\nud{{\nu {\rm d}}}

\def\SBBN{{\rm (0)}}

\def\Yp{Y_{\rm p}}

\graphicspath{Figures_v1/}

\begin{document}

\title{Consistent $N_{\rm eff}$ fitting in big bang nucleosynthesis analysis}

\author{Sougata Ganguly}
\email{sganguly0205@ibs.re.kr}
\affiliation{Particle Theory  and Cosmology Group (PTC),
Center for Theoretical Physics of the Universe (CTPU), \\
Institute for Basic Science, Daejeon 34126, Republic of Korea}

\author{Tae Hyun Jung}
\email{thjung0720@gmail.com}
\affiliation{Particle Theory  and Cosmology Group (PTC),
Center for Theoretical Physics of the Universe (CTPU), \\
Institute for Basic Science, Daejeon 34126, Republic of Korea}

\author{Seokhoon Yun}
\email{seokhoon.yun@knu.ac.kr}
\affiliation{Department of Physics, Kyungpook National University, Daegu 41566, Korea}
\affiliation{Particle Theory  and Cosmology Group (PTC),
Center for Theoretical Physics of the Universe (CTPU), \\
Institute for Basic Science, Daejeon 34126, Republic of Korea}

\preprint{CTPU-PTC-25-29}

\begin{abstract}
The effective number of neutrino species, $N_{\rm eff}$, serves as a key fitting parameter extensively employed in cosmological studies.
In this work, we point out a fundamental inconsistency in the conventional treatment of $N_{\rm eff}$ in big bang nucleosynthesis (BBN), particularly regarding its applicability to new physics scenarios where $\Delta N_{\rm eff}$, the deviation of $N_{\rm eff}$ from the standard BBN prediction, is negative.
To ensure consistent interpretation, it is imperative to either restrict the allowed range of $N_{\rm eff}$ or systematically adjust neutrino-induced reaction rates based on physically motivated assumptions. 
As a concrete example, we consider a simple scenario in which a negative $\Delta N_{\rm eff}$ arises from entropy injection into the electromagnetic sector due to the decay of long-lived particles after neutrino decoupling.
This process dilutes the neutrino density and suppresses the rate of neutrino-driven neutron-proton conversion.
Under this assumption, we demonstrate that the resulting BBN constraints on $N_{\rm eff}$ deviate significantly from those obtained by the conventional, but unphysical, extrapolation of dark radiation scenarios into the $\Delta N_{\rm eff} < 0$ regime.
\end{abstract}

\maketitle

\noindent
\section{Introduction}

The success of big bang nucleosynthesis (BBN) in predicting the primordial abundances of $\ce{^4He}$, $\D$, and $\ce{^3He}$ has provided strong constraints on new physics 
(see Refs.~\cite{Pitrou:2018cgg,Fields:2019pfx} for reviews).
While detailed case-by-case analysis is essential, a model-independent approach has been frequently achieved through fitting in terms of effective parameters.
Obviously, the most important among them is the baryon asymmetry parameter $\eta_B\equiv n_B/n_\gamma$, which is defined as the ratio of the baryon number density $n_B$ and the photon number density $n_\gamma$.
It determines the deuterium bottleneck temperature and the overall scale of nuclear reaction rates, which in turn affect both the initiation and finalization of BBN. 
Since the BBN procedure is insensitive to the details of how the baryon asymmetry is generated, $\eta_B$ provides a convenient and robust way to characterize the baryonic input to BBN across various new physics scenarios.

Another widely-used parameter is the effective number of neutrino species $N_{\rm eff}$, which is introduced to parametrize the energy density of radiation other than photons, and its consequence in the Hubble rate.
In the standard BBN scenario, neutrinos decouple from the thermal plasma slightly before electron-positron annihilation.
After decoupling, the $e^\pm$ annihilation heats the photon bath but not the neutrinos, resulting in the well-known temperature ratio $T_\nu/T_\gamma = (4/11)^{1/3}$.
This sets the definition of 
\bal
N_{\rm eff} \equiv \frac{\rho_{\rm rad}-\rho_{\gamma}}
{
\frac{7}{8} \left(\frac{4}{11}\right)^{4/3} \rho_\gamma
}
\, ,
\eal
where the photon energy density is given by $\rho_\gamma = 2 \cdot \pi^2 T_\gamma^4/30$, and $\rho_{\rm rad}$ is the total radiation energy density
(although the plasma temperature $T$ is conventionally denoted for the photon temperature, we also denote it as $T_\gamma$ for clarity).
Since $\rho_{\rm rad}-\rho_\gamma=\rho_\nu$ in the standard scenario and there are three neutrino species, $N_{\rm eff} \simeq 3$ is expected.
Incorporating slight corrections from non-instantaneous decoupling and finite temperature QED effects, more precise prediction is given by $N_{\rm eff}^{(0)} \simeq 3.045$~\cite{Mangano:2001iu, Mangano:2005cc, deSalas:2016ztq, EscuderoAbenza:2020cmq, Akita:2020szl, Cielo:2023bqp}.

$N_{\rm eff}$ affects the evolution of the Hubble parameter as
\bal
H(T_\gamma) = H^\SBBN(T_\gamma)\sqrt{1+\frac{\rho_\nu^\SBBN(T_\gamma)}{\rho_{\rm tot}^\SBBN(T_\gamma)} \frac{\Delta N_{\rm eff}}{N_{\rm eff}^{(0)}}}
\, ,
\label{Eq:HinNeff}
\eal
where $\Delta N_{\rm eff} \equiv N_{\rm eff}-N_{\rm eff}^{\SBBN}$, $\rho_{\rm tot}(T_\gamma)$ and $\rho_{\nu}(T_\gamma)$ are energy densities of the whole plasma and the neutrino sector at a photon temperature $T_\gamma$, $H(T_\gamma)$ is the Hubble rate, and the superscript $\SBBN$ indicates these quantities evaluated in the standard scenario.
The modified Hubble rate for a nonzero $\Delta N_{\rm eff}$ is well-known to change the time scale in the BBN era; enhanced $H$ makes overall reaction rates less efficient.
This is how the observed primordial abundances of light elements could constrain $N_{\rm eff}$, conventionally.

In this paper, however, we question whether $N_{\rm eff}$ is indeed a suitable effective parameter in BBN.
Does it have an obvious physical interpretation across a wide range of new physics models?
When $\Delta N_{\rm eff}$ is positive, the answer is clearly yes, as it represents dark radiation that just modifies the Hubble expansion rate.
Properties other than the energy density of the dark radiation do not change the BBN process as long as it does not decay and is decoupled from the visible sector long before the BBN era.
Many new physics models containing feebly interacting light particles are constrained by the same upper bound of $N_{\rm eff}$ once it is obtained.

However, when $\Delta N_{\rm eff} < 0$, its physical interpretation becomes less clear, as it cannot be explained by additional relic (because energy density cannot be negative). 
Instead, a negative $\Delta N_{\rm eff}$ \emph{inevitably} requires modifications in the relation between photon and neutrino temperatures, and thus the neutrino-induced $n\leftrightarrow p$ conversion rates must be modified accordingly.
However, the corresponding correction has not been considered in most literature (see, for instance, fitting for the BBN conducted in Refs.~\cite{Boehm:2013jpa,Vogel:2013raa, Buen-Abad:2015ova, Chacko:2015noa, Kelly:2020aks,Giovanetti:2021izc,Adshead:2022ovo}).
A lower bound of $\Delta N_{\rm eff}$ obtained without this correction cannot be applied to \emph{any} new physics model even if they predict a negative $\Delta N_{\rm eff}$.

\begin{figure}[t]
    \centering
    \includegraphics[width=0.45\textwidth]{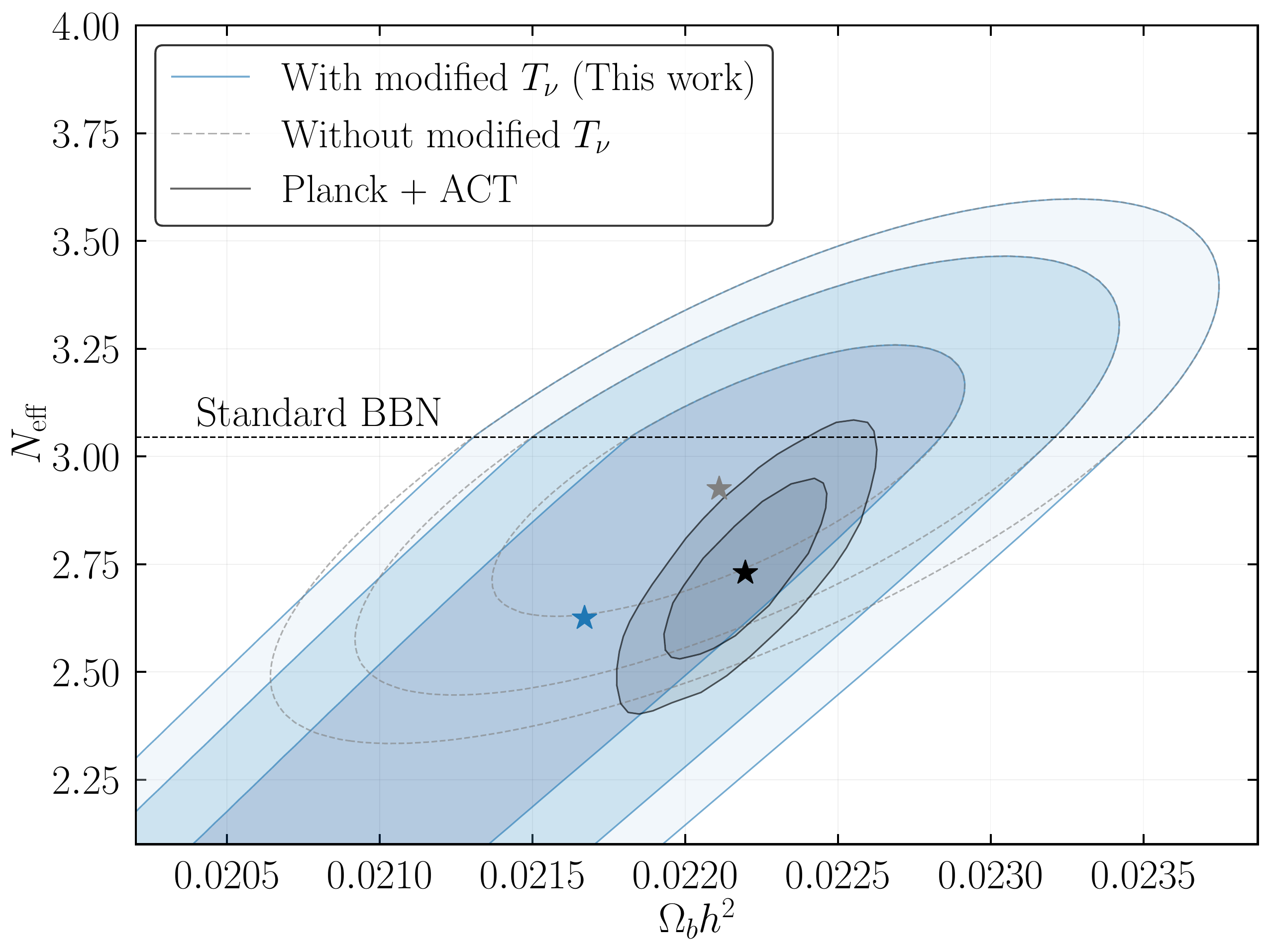}
    \caption{Our fitting result in
    $\Omega_b h^2$\,--\,$N_{\rm eff}$ plane
    using the PDG recommended values of
    $\Yp$ and $\rm D/H$. Blue shaded region
    outlined by the blue solid line depicts
    our result considering modified $T_\nu$
    for $\Delta N_{\rm eff} < 0$.  The dashed contours denote the fitting result in the conventional scenario, and the prediction of Planck+ACT is denoted by the gray shaded regions.
    The SBBN value of $N_{\rm eff}$ is depicted by the horizontal dashed line.
    }
    \label{fig:summary}
\end{figure}

Before we present the details of our study, let us clarify our conclusion to avoid any misinterpretation of our results: 
\begin{itemize}
    \item[1.]{$N_{\rm eff}$ is not a good fitting parameter in the BBN analysis. 
    We recommend not to use $N_{\rm eff}$ in BBN analysis.}
    \item[2.]{If it is insisted to use $N_{\rm eff}$, some model dependence is inevitable. 
    One may simply include minimal corrections in the neutrino-induced $n\leftrightarrow p$ conversion rates for $\Delta N_{\rm eff} < 0$, but it does not give a meaningful result in the end, as shown in Fig.~\ref{fig:summary}.}
\end{itemize}
Fig.~\ref{fig:summary} shows our fitting result in the $\Omega_b h^2$
\,--\,$N_{\rm eff}$ plane obtained with the simplest assumption on neutrino temperature for the $\Delta N_{\rm eff}<0$ region; we assume that the negative $\Delta N_{\rm eff}$ solely comes from the change in $T_\nu/T_\gamma$ (the details of the assumptions are given in the next section, and the result shown in Fig.\,\ref{fig:summary} must be applied only to models/scenarios that satisfy our assumptions).
The blue solid lines are contours corresponding to $68\,\%$, $95\,\%$ and $99\,\%$ confidence level (C.L.) that we obtain by including the modified neutrino-to-photon temperature ratio at $\Delta N_{\rm eff}<0$.
Here, we adopt the observed values of the primordial $^4\He$ mass fraction $\Yp\equiv \rho(^4\He)/\rho_b$ and the deuterium-to-hydrogen ratio $\D/\H$ as
\bal
&\Yp^{\rm (obs)} = 0.245 \pm 0.003,  \label{Eq:Yp_obs}
\\
&(\D/\H)^{\rm (obs)} = (2.547 \pm 0.029)\times 10^{-5},
\eal
as recommended by particle data group (PDG)\,\cite{ParticleDataGroup:2022pth}. 

As shown in the figure, BBN is not useful to estimate the lower bound of $N_{\rm eff}$.
The lower edges of the BBN contours do not even appear, while the black solid contours—representing the 68\,\% and 95\,\% C.L. from the CMB analysis in Fig.\,2 of Ref.\,\cite{ACT:2025tim} (which combines the ACT and Planck datasets\,\cite{Planck:2018vyg})—are much narrower than the BBN contours.
This happens because $^4{\rm He}$ abundance is insensitive to both $N_{\rm eff}$ and $\eta_B$ for $\Delta N_{\rm eff}<0$, where we find an accidental cancellation in $^4{\rm He}$ abundance among the effects coming from reduced neutrino temperature, reduced Hubble rate, and a shifted equilibrium of the neutron fraction.
Therefore, the $^4{\rm He}$ abundance does not give a useful constraint on the $\Delta N_{\rm eff}<0$ region.
Then, we have only one constraint coming from the ${\rm D}$ abundance, making the diagonal direction in Fig.\,\ref{fig:summary} observationally indistinguishable in BBN.

The rest of this paper is organized as follows.
In Section\,\ref{sec:assumptions}, we explain our assumptions and the motivations for why we assume them.
In Section\,\ref{sec:np_conversion}, we summarize the modified neutrino-induced reaction rates, and how they affect the primordial abundances.
We show our results in Section\,\ref{sec:results}, and conclude in Section\,\ref{sec:discussions}.

\section{
Assumptions for exploring $\Delta N_{\rm eff}$ in BBN}
\label{sec:assumptions}

To explore the effects of varying $\Delta N_{\rm eff}$ during BBN, we adopt a minimal set of assumptions that allow for a model-independent interpretation within two distinct regimes: $\Delta N_{\rm eff} > 0$ and $\Delta N_{\rm eff} < 0$.

For $\Delta N_{\rm eff} > 0$, we assume that the neutrino sector remains unaltered from the standard scenario, and the additional energy density arises solely from a dark radiation component which decouples from the Standard Model (SM) plasma well before neutrino decoupling and remains relativistic throughout the BBN epoch.
This assumption is typical in the literature.
Since it does not interact with the SM plasma, it does not affect either the neutrino or photon temperatures, and the standard neutrino-to-photon temperature ratio remains unchanged as follows
\bal
\frac{T_{\nu_e}(T_\gamma)}{T_{\nu_e}^\SBBN\!(T_\gamma)}
=
1  \quad \text{for $\Delta N_{\rm eff}>0$},
\label{Eq:Tnu_ratio_DR}
\eal
where $T_{\nu_e}(T_\gamma)$ denotes the modified electron neutrino temperature as a function of photon temperature $T_\gamma$, and $T_{\nu_e}^\SBBN(T_\gamma)$ is the corresponding temperature in the standard scenario. 
During the BBN era, both neutrinos and dark radiation redshift as $a^{-4}$, where $a$ is the scale factor, so their energy density ratio remains constant.
This allows us to express the energy density of dark radiation as $\rho_{\rm DR} = (\Delta N_{\rm eff}/N_{\rm eff}) \rho_\nu$, justifying the use of Eq.~\eqref{Eq:HinNeff} for the modified Hubble rate.

For $\Delta N_{\rm eff} < 0$, we assume that the deviation arises entirely from a relatively suppressed neutrino temperature, i.e., a reduced $T_{\nu_e}/T_\gamma$ ratio.
This is the simplest viable assumption, since a negative $\Delta N_{\rm eff}$ necessarily implies a deviation in the neutrino-to-photon temperature ratio.
The resulting Hubble rate is again given by Eq.~\eqref{Eq:HinNeff}.

Such a situation can be realized in scenarios where a dark-sector particle, denoted by $\chi$, decays electromagnetically\footnote{
If $\chi$ decays into hadrons, the neutron-to-proton ratio can be significantly altered by injected charged pions, kaons, and baryons, necessitating additional assumptions about the decay branching ratios~\cite{Reno:1987qw, Kohri:1999ex, Kohri:2001jx, Pospelov:2010cw, Jung:2025dyo}.
} after the neutrino decoupling (at $T_{\nu{\rm d}}\simeq 2\,\MeV$) but before the neutron-proton freeze-out (at $T_{np}\simeq 0.8\,\MeV$).
If the decay occurs at $T_\gamma > T_{\nu{\rm d}}$, the injected entropy is shared with the neutrino sector, thereby preserving the standard thermal scenario.
On the other hand, if the decay occurs at $T_\gamma < T_{np}$, the analysis becomes more complicated because one must track the time dependence of the neutrino temperature ratio as well as the effect of $\chi$ on the early-time Hubble rate.

To avoid this complication, we focus on the case in which the decay of $\chi$ occurs effectively instantaneously in the interval $T_{np} < T_\gamma < T_{\nu{\rm d}}$, which enables a simplified BBN analysis with $N_{\rm eff}$.
Even in this case, the treatment should be translated as an approximation.
Nevertheless, it is sufficient for predicting the light-element abundances, since the neutron-to-proton ratio freezes out near $T_{np}$ and is not sensitive to the earlier thermal history.

Furthermore, we do not include scenarios of early matter domination driven by $\chi$.\footnote{
This restricts $m_\chi \lesssim 1\,{\rm GeV} \left(\dfrac{6.45\times 10^{-4}}{Y^{(0)}_\chi}\right)
\left(\dfrac{1\,\rm sec}{\tau_\chi}\right)^{1/2}$
where $m_\chi$, $Y_\chi^{(0)}$, and $\tau_\chi$
are the mass, initial yield, and lifetime of
$\chi$, respectively.
}
In such a case, the difference between the kinetic and chemical decoupling of neutrinos induces an overall suppression factor to be multiplied in the neutrino distribution function, so we need an additional parameter.
For the detailed studies, see Refs.~\cite{Kawasaki:1999na, Kawasaki:2000en, Hasegawa:2019jsa, Hasegawa:2020ctq}, where the deformation from the thermal equilibrium distribution and the effects of neutrino oscillation were also carefully studied.

For simplicity and model independence, we adopt a baseline parametrization in which the modification of the electron-neutrino temperature is written as
\bal
\frac{T_{\nu_e}(T_\gamma)}{T_{\nu_e}^\SBBN\!(T_\gamma)}
=
\left(1 + \frac{x_e \Delta N_{\rm eff}}{N_{\rm eff}^\SBBN}\right)^{1/4}\,.
\label{Eq:Tnu_ratio}
\eal
Here, $x_e$ parametrizes the fact that $\nu_e$ decouples slightly later than $\nu_x$ ($x=\mu,\tau$), and can therefore be diluted less efficiently by entropy injection around the decoupling epoch.
The flavor-universal limit corresponds to $x_e=1$, in which case all neutrino flavors are diluted equally, $\frac{T_{\nu_e}(T_\gamma)}{T_{\nu_e}^\SBBN\!(T_\gamma)}=\frac{T_{\nu_\mu}(T_\gamma)}{T_{\nu_\mu}^\SBBN\!(T_\gamma)}=\frac{T_{\nu_\tau}(T_\gamma)}{T_{\nu_\tau}^\SBBN\!(T_\gamma)}$
and the above expression reduces to the simplest baseline treatment.

In the following analysis, we take this baseline treatment in Eq.~\eqref{Eq:Tnu_ratio}, starting from a large temperature.
It means that in our analysis, the neutrino temperature is already shifted before the neutrino decoupling, which is unrealistic.
However, the final neutron-to-proton ratio is only determined around $T_{np}$ at which Eq.~\eqref{Eq:Tnu_ratio} is valid, and thus our treatment provides a numerically valid result.

In Appendix.~\ref{App:Refined}, we examine conditions on $\chi$'s lifetime $\tau_\chi$ (or its decay temperature $T_{\rm decay}$ at which $\tau_\chi^{-1} = 3H(T_{\rm decay})$) under which our simplified treatment agrees with a more refined calculation where the entropy injection is treated continuously rather than instantaneously.
Indeed, we find that the baseline treatment using Eq.~\eqref{Eq:Tnu_ratio} with $x_e = 1$ remains a good approximation when $0.1\,\sec \lsim \tau_\chi \lsim 1\,\sec$ (i.e., $T_{\nu{\rm d}} \gtrsim T_{\rm decay} \gtrsim T_{np} $).
However, for $\tau_\chi \lesssim 0.1\,\sec$, the effect of $x_e\neq1$ becomes important, in which case a more model-dependent analysis is required; see, e.g., Refs.~\cite{Ibe:2019gpv, Li:2020roy, Yeh:2024ors}.
For longer lifetimes, as long as the entropy injection still occurs before neutron-to-proton freeze-out (i.e., $T_{\rm decay} > T_{np}$), the baseline treatment remains in good agreement with the numerical results.
Interestingly, this agreement persists even up to $\tau_\chi \sim 100\,\sec$, corresponding to the epoch of the deuterium bottleneck.
For $\tau_\chi \gtrsim 100\,\sec$, however, the dominant additional effect relevant for BBN is the dilution of the baryon asymmetry, which must be included in order to evaluate the deuterium abundance correctly; see Appendix.~\ref{App:Refined} for details.

Similarly, constraints derived under the dark radiation assumption for $\Delta N_{\rm eff} > 0$ cannot be applied to models in which the excess energy is deposited into the neutrino sector (see, e.g., Refs.~\cite{Escudero:2019gzq,Sabti:2019mhn,Esseili:2023ldf,Chang:2024mvg,Deppisch:2024izn,Kanzaki:2007pd}).
In such cases, a more precise treatment of the neutrino momentum distribution is required, and $\Delta N_{\rm eff}$ alone is not sufficient to characterize the scenario.
In fact, it is even possible to have a small $\Delta N_{\rm eff}$ due to cancellations between competing dark sector effects, despite a significant shift in $T_\nu/T_\gamma$~\cite{Hong:2023fcy}.

These limitations underscore that $N_{\rm eff}$ may not serve as a universally reliable fitting parameter for BBN analyses. 
Nevertheless, in the remainder of this work, we perform BBN fits using Eqs.~\eqref{Eq:Tnu_ratio_DR} and \eqref{Eq:Tnu_ratio}, each of which is consistent with certain classes of new physics scenarios. 
It is important to emphasize that conventional extrapolations of Eq.~\eqref{Eq:Tnu_ratio_DR} into the $\Delta N_{\rm eff} < 0$ region do not correspond to any physically realizable model and thus yield unphysical results. 
Likewise, extrapolating Eq.~\eqref{Eq:Tnu_ratio} into the $\Delta N_{\rm eff} > 0$ regime does not produce accurate predictions, as the momentum distribution of injected neutrinos becomes important in that case.

To avoid such inconsistencies, we separately constrain $\Delta N_{\rm eff}$ using Eq.~\eqref{Eq:Tnu_ratio_DR} for $\Delta N_{\rm eff}>0$ and Eq.~\eqref{Eq:Tnu_ratio} for $\Delta N_{\rm eff}<0$.
Since the two are continuously connected at $\Delta N_{\rm eff}=0$, we combine the results into a single plot, as shown in Fig.~\ref{fig:summary}.

\section{Impact on Neutron-Proton Freeze-Out}
\label{sec:np_conversion}

\subsection{Weak interaction rates}

We now examine how a nonzero shift in $\Delta N_{\rm eff}$ modifies the neutron-proton freeze-out process by altering both the Hubble rate $H$ and the neutrino temperature $T_\nu$.
The evolution of the neutron-to-baryon ratio, defined as $X_n \equiv n_n / n_B$, where $n_n$ is the neutron number density, is governed by the Boltzmann equation
\bal
&\frac{dX_n}{dt} 
\! = \!
-\Gamma_{n\to p}(T_\gamma, T_\nu) X_n \!
+\Gamma_{p\to n}(T_\gamma, T_\nu) X_p \!
+\cdots ,
\eal
where $X_p \simeq 1 - X_n$ denotes the proton-to-baryon ratio, and the ellipsis represents contributions from other nuclear reactions, which are negligible in the temperature range $T_\gamma > T_{np}$.
The rates $\Gamma_{n\to p}$ and $\Gamma_{p\to n}$ correspond to neutron-to-proton and proton-to-neutron conversions via weak interactions, respectively.

When neutrinos/anti-neutrinos and electrons/positrons follow equilibrium distributions, the individual reaction rates contributing to $\Gamma_{n\to p}$ and $\Gamma_{p\to n}$ can be written in analytic form~\cite{Kolb:1990vq, Weinberg:2008zzc, Mukhanov:2003xs}
\bal
&\Gamma_{n\nu_e\to pe^-}
= \frac{1+3g_A^2}{2\pi^3}G_F^2 Q^5 J_{1}^\infty(T_\gamma,T_{\nu_e})\,,
\label{Eq:Gamma_1}
\\
&\Gamma_{n e^+\to p\bar\nu_e}
= \frac{1+3g_A^2}{2\pi^3}G_F^2 Q^5 J_{-\infty}^{-m_e/Q}(T_\gamma,T_{\nu_e})\,,
\label{Eq:Gamma_2}
\\
&\Gamma_{p\bar\nu_e\to n e^+}
= \frac{1+3g_A^2}{2\pi^3}G_F^2 Q^5 K_{-\infty}^{-m_e/Q}(T_\gamma,T_{\nu_e})\,,
\label{Eq:Gamma_3}
\\
&\Gamma_{pe^-\to n\nu_e}
= \frac{1+3g_A^2}{2\pi^3}G_F^2 Q^5 K_{1}^\infty(T_\gamma,T_{\nu_e})\, ,
\label{Eq:Gamma_4}
\eal
where $g_A \simeq 1.27$, $G_F$ is the Fermi constant, and $Q \equiv m_n - m_p \simeq 1.293\,\MeV$ is the neutron-proton mass difference.
The $J_a^b$ and $K_a^b$ functions are defined by
\bal
&J_a^b(T_\gamma,T_{\nu_e})= \!\!
\int_a^b \!\!
\sqrt{1-\frac{(m_e/Q)^2}{q^2}}
\frac{q^2(q-1)^2 dq}
{(1+e^{-\frac{Q}{T_\gamma}q})(1+e^{\frac{Q}{T_{\nu_e}}(q-1)})}\,,
\\
&K_a^b(T_\gamma,T_{\nu_e})= \!\!
\int_a^b \!\!
\sqrt{1-\frac{(m_e/Q)^2}{q^2}}
\frac{q^2(q-1)^2 dq}
{(1+e^{\frac{Q}{T_\gamma}q})(1+e^{-\frac{Q}{T_{\nu_e}}(q-1)})}\,.
\eal
We also include the neutron decay rate $\Gamma_n = (878.4\,\sec)^{-1}$~\cite{ParticleDataGroup:2024cfk} as part of the total $\Gamma_{n\to p}$.

Before freeze-out, when $\Gamma_{n\to p}\,,\Gamma_{p\to n} \gg H$, the neutron-to-baryon ratio follows its equilibrium value
\bal
X_{n,{\rm fo}}(T, T_{\nu_e}) \simeq \frac{\Gamma_{p\to n}(T,T_{\nu_e})}{\Gamma_{n\to p}(T,T_{\nu_e})+\Gamma_{p\to n}(T,T_{\nu_e})} \, .
\label{Eq:Xn_eq}
\eal
Note that $T_{\nu_e}$ is a function of $T_\gamma$.
In the standard BBN (SBBN) scenario, the freeze-out temperature $T_{np}^\SBBN$ is close to or slightly higher than the electron mass, $m_e \simeq 0.511\,\MeV$, so the neutrino temperature is approximately equivalent to the photon temperature, $T_{\nu_e}^\SBBN \simeq T_\gamma$.
This leads to the standard relations of $\Gamma^{(0)}_{p\to n}(T,T) \simeq \Gamma^{(0)}_{n\to p}(T,T) e^{-Q/T}$, which yield a freeze-out value of $X_n \approx 0.17$.

At the deuterium bottleneck temperature $T_\D$, almost all neutrons become $^4\He$.
Thus, $X_n$ is directly related to the observable $\Yp \simeq 2 X_n|_{T=T_\D} \simeq 2 X_{n,{\rm fo}}|_{T=T_{np}} e^{-\Gamma_n (t_\D-t_{np})}$, where the exponential factor comes from the fact that $X_n$ slightly decreases from its freeze-out value due to the beta decay of the neutron.
This allows us to approximately consider 
\bal
\frac{\delta \Yp}{\Yp^\SBBN} \simeq \frac{\delta X_{n,{\rm fo}} \big( T_{np}, \, T_{\nu_e}(T_{np}) \big)}{X_{n,{\rm fo}} \big(T_{np}^\SBBN, \, T_{\nu_e}^\SBBN \!(T_{np}^\SBBN) \big)},
\label{Eq:deltaYpoverYp0}
\eal
where $\Yp^\SBBN$ is the $\Yp$ predicted in the SBBN, and $\delta \Yp \equiv \Yp-\Yp^\SBBN$ is the deviation from the SBBN for a nonzero $\Delta N_{\rm eff}$ case.
For simplicity, we denote the right-hand-side of Eq.\,\eqref{Eq:deltaYpoverYp0} as simply $\delta X_{n,{\rm fo}}/X_{n,{\rm fo}}^{(0)}$.

In the presence of additional dark radiation (i.e., $\Delta N_{\rm eff} > 0$ with $T_{\nu_e}/T_{\nu_e}^\SBBN = 1$), the increased Hubble expansion rate reduces the efficiency of the weak interactions, as $\frac{d}{dt} \simeq - H \frac{d}{d\ln T}$.
Consequently, the freeze-out occurs earlier ($T_{np}<T_{np}^\SBBN$), when $X_n$ is higher, leading to a larger final abundance of $\ce{^4 He}$.

For the case of $\Delta N_{\rm eff} < 0$, where entropy is injected into the visible sector, the $n \leftrightarrow p$ conversion rates are further modified due to the reduced neutrino temperature; see Eq.~\eqref{Eq:Tnu_ratio}.
This alters $X_n^{\rm (eq)}$ as well as $T_{np}$, since both $H$ and the weak rates are modified.
Combining these effects, we find that the freeze-out value of $X_n$ decreases as $N_{\rm eff}$ decreases, though the sensitivity is relatively mild.

\subsection{Analytic Estimates of $X_n$ Sensitivity to $\Delta N_{\rm eff}$}
\label{sec:analytic}

While exact numerical results are obtained separately using full calculations, we now analyze how small deviations in $N_{\rm eff}$ affect the neutron-to-baryon ratio $X_n$, focusing on linear-order corrections.
This analysis is intended to capture the qualitative parametric dependence.

The leading-order modification to the Hubble rate can be expressed as
\be
H \simeq  H^\SBBN\left(1 + \frac{7}{86}\Delta N_{\rm eff}\right) \,.
\ee
On the other hand, the weak reaction rates depend on $T_{\nu_e}$, while we have different $T_{\nu_e}$ for $\Delta N_{\rm eff} > 0$ and $\Delta N_{\rm eff} < 0$.
These different relations can be combined by Eq.\,\eqref{Eq:Tnu_ratio}; we take $x_e=0$ for $\Delta N_{\rm eff}>0$ and $x_e=1$ for $\Delta N_{\rm eff}<0$.
To linear order, it gives
\be
\frac{\delta T_{\nu_e}}{T_{\nu_e}^\SBBN} \simeq \frac{x_e}{12}\Delta N_{\rm eff} \,.
\ee

Since the neutron-proton conversion rates are sensitive to the electron-neutrino temperature, we examine how small perturbations in $T_{\nu_e}$ affect them.
At the temperatures of interest, the relevant rates approximately scale as 
\bal
&\Gamma_{n \nu_e \to p e^-}(T_\gamma, T_{\nu_e}) 
\propto 
\left(\frac{T_{\nu_e}}{Q}\right)^5 
+ \frac{1}{2}\left(\frac{T_{\nu_e}}{Q}\right)^4
\\
&\Gamma_{p \bar{\nu}_e \to n e^+} (T_\gamma, {T_{\nu_e}})
\propto 
\left(\frac{T_{\nu_e}}{Q}\right)^4 \exp\left[-\frac{Q + m_e}{T_{\nu_e}}\right]
\\
&\Gamma_{n e^+ \to p \bar \nu_e} (T_\gamma, T_{\nu_e})
\propto 
\left(\frac{T_\gamma}{Q}\right)^5 
+ \frac{1}{2}\left(\frac{T_\gamma}{Q}\right)^4
\\
&\Gamma_{p e^- \to n \nu_e} (T_\gamma, {T_{\nu_e}})
\propto 
\left(\frac{T_\gamma}{Q}\right)^5 \exp\left[-\frac{Q - m_e}{T_\gamma}\right]
\eal
Thus, to linear order in $\delta T_{\nu_e}/T_{\nu_e}^\SBBN$, we find
\bal
\dfrac{\Gamma_{n \to p}(T_\gamma, T_{\nu_e})}{\Gamma_{n \to p}(T_\gamma, T_{\nu_e}^\SBBN)}
&\simeq 1  
+ \left(2 + \dfrac{T_{\nu_e}^{(0)}}{Q + 2 T_{\nu_e}^{(0)}}\right)
\dfrac{\delta T_{\nu_e}}{T_{\nu_e}^{(0)}} 
\,,
\\
\dfrac{\Gamma_{p \to n}(T_\gamma, T_{\nu_e})}{\Gamma_{p \to n}(T_\gamma, T_{\nu_e}^\SBBN)}
&\simeq 1 
+ \left( 2 + \dfrac{Q+m_e}{2T_{\nu_e}^{(0)}}\right)
\dfrac{\delta T_{\nu_e}}{T_{\nu_e}^{(0)}}
\, .
\eal 
Recall that $T_{\nu_e}$ and $T_{\nu_e}^\SBBN$ are approximately linear functions of $T_\gamma$, while $\delta T_\nu/T_\nu^\SBBN$ is constant.

The $n/p$ freeze-out temperature $T_{np}$ is determined by the condition
\bal
\Big[ \Gamma_{n \to p} + \Gamma_{p \to n} \Big]_{T=T_{np}} \simeq H (T_{np})\,.
\eal
We take the SBBN value $T_{np}^{\SBBN} \simeq 0.82\,\MeV$, which gives the equilibrium abundance $X_{n, \rm fo}^\SBBN = \frac{e^{-Q/T_{np}^\SBBN}}{1 + e^{-Q/T_{np}^\SBBN}} \simeq 0.17$.
Around $T_{np}^{\SBBN}$, we can also approximate $T_\nu^{(0)} \simeq T_\gamma$, and thus $\Gamma_{n\to p}(T_{ np}^\SBBN, T_{np}^\SBBN) \simeq \frac{1}{1+e^{-Q/T_{np}^{\SBBN}}}H(T_{np}^\SBBN)$ and $\Gamma_{p\to n}(T_{ np}^\SBBN, T_{np}^\SBBN) \simeq \frac{e^{-Q/T_{np}^{\SBBN}}}{1+e^{-Q/T_{np}^{\SBBN}}}H(T_{np}^\SBBN)$.
Linearizing in $\delta T_\nu$, we obtain the shift in the freeze-out temperature, $\delta T_{np} \equiv T_{np} - T_{np}^{(0)}$,
\bal
\dfrac{\delta T_{np}}{T_{np}^{(0)}}  
&\simeq (0.03 - 0.07 x_e) \Delta N_{\rm eff}\,.\label{Eq:deltaTnp_EI}
\eal
Consequently, the freeze-out value of the neutron fraction becomes
\bal
\frac{\delta X_{n,{\rm fo}} }{X_{n,{\rm fo}}^{(0)}}
&\simeq  (0.04 - 0.04 x_e) \Delta N_{\rm eff} \nn \\
& \simeq {\cal O}(10^{-3}) \Delta N_{\rm eff} \quad \text{for } x_e = 1
\,.\label{Eq:Xnfo_EI}
\eal
The coefficient of $\Delta N_{\rm eff}$ in Eq.\,\eqref{Eq:Xnfo_EI} is an order of magnitude smaller than the individual coefficients
of $\Delta N_{\rm eff}$ in Eq.\,\eqref{Eq:deltaTnp_EI}.
This implies that there exists a cancellation at $10\,\%$-tuning level among the different effects such as shifts of $T_{\nu_e}$, $H$, and $X_{n,{\rm fo}}$.
Since there is no symmetry that guarantees this cancellation, we consider it accidental.

The shift in the freeze-out temperature
for the dark radiation case ($\Delta N_{\rm eff} > 0$) can be obtained
by putting $x_e = 0$ in Eq.\,\eqref{Eq:deltaTnp_EI}, and it is given by
\bal
\dfrac{\delta T_{np}}{T_{np}^{(0)}} \simeq 0.03 \Delta N_{\rm eff}\,,
\label{eq:deltaTnp_DR}
\eal
and hence,
\bal
\frac{\delta X_{n,{\rm fo}} }{X_{n,{\rm fo}}^{(0)}}
 = 0.04\Delta N_{\rm eff}\,\,.
 \label{Eq:Xnfo_DR}
\eal
These results, Eqs.~\eqref{Eq:Xnfo_EI} and \eqref{Eq:Xnfo_DR}, demonstrate that the sensitivity of $X_n$ to $\Delta N_{\rm eff}$ is one order of magnitude weaker in the entropy injection case ($\Delta N_{\rm eff} < 0$).
This suppression arises from an accidental cancellation between the competing effects of the reduced equilibrium abundance and shifts in $T_{np}$ and $X_n^{\rm (eq)}$.
As a result, setting a robust lower bound on $\Delta N_{\rm eff}$ using BBN becomes challenging under electromagnetic injection scenarios.

\subsection{Numerical implementation with \texttt{PArthENoPE}}
For a quantitatively more precise study,
we perform numerical calculations using the publicly available \texttt{PArthENoPE} 3.0 code~\cite{Pisanti:2007hk, Consiglio:2017pot, Gariazzo:2021iiu}, incorporating appropriate modifications to the weak interaction rates.
The default implementation of weak interaction rates in \texttt{PArthENoPE} includes higher-order corrections~\cite{Serpico:2004gx}, which are beyond the scope of our present analysis.
Instead, we restrict ourselves to the leading-order expressions for the neutron-proton conversion rates, as defined in Eqs.~\eqref{Eq:Gamma_1}-\eqref{Eq:Gamma_4}.
To implement the effects of $\delta T_\nu$, we define correction factors as
\bal
x_{np} = \frac{\Gamma_{n\to p}}{\Gamma_{n\to p}^{(0)}} 
\,, \quad
x_{pn} = \frac{\Gamma_{p\to n}}{\Gamma_{p\to n}^{(0)}}\,,
\label{Eq:xnp_xpn}
\eal
where $\Gamma_{n\to p}^{(0)}$ and $\Gamma_{p\to n}^{(0)}$ denote the standard leading-order rates.
We then rescale the weak interaction rates in the code by $x_{np}$ and $x_{pn}$, respectively.
This approach consistently incorporates the effects of a modified neutrino temperature up to the leading order, which is sufficient for our purposes in this study.

In the \texttt{PArthENoPE} framework, we access the standard neutrino-to-photon temperature ratio $T_\nu^\SBBN/T_\gamma$ which is defined as
\bal
\frac{T_\nu^\SBBN}{T_\gamma} \equiv  \left(\frac{\hat{\rho}_{e\gamma B}(T_\gamma)+\hat{p}_{e\gamma B}(T_\gamma)}{\hat{\rho}_{e\gamma B}(T_\nud)+\hat{p}_{e\gamma B}(T_\nud)}\right)^{1/3} \,,
\eal
for photon temperature $T_\gamma < T_\nud \equiv 2.3\,\MeV$.
Then, we use Eqs.~\eqref{Eq:Tnu_ratio_DR} and \eqref{Eq:Tnu_ratio} to obtain $T_\nu$ for a given $N_{\rm eff}$ input, and calculate the correction factors $x_{np}$ and $x_{pn}$ at each time step.
Here, $\hat{\rho}_{e\gamma B}(T) \equiv \rho_{e\gamma B}/T^4$ and $\hat{p}_{e\gamma B}(T) \equiv p_{e\gamma B}/T^4$, with $\rho_{e\gamma B}$ and $p_{e\gamma B}$ denoting the energy density and pressure of the electron-photon-baryon plasma, respectively.
For $T_\gamma > T_\nud$, the temperature ratio is fixed to unity.

\begin{figure*}
    \centering
    \includegraphics[width=0.48\textwidth]{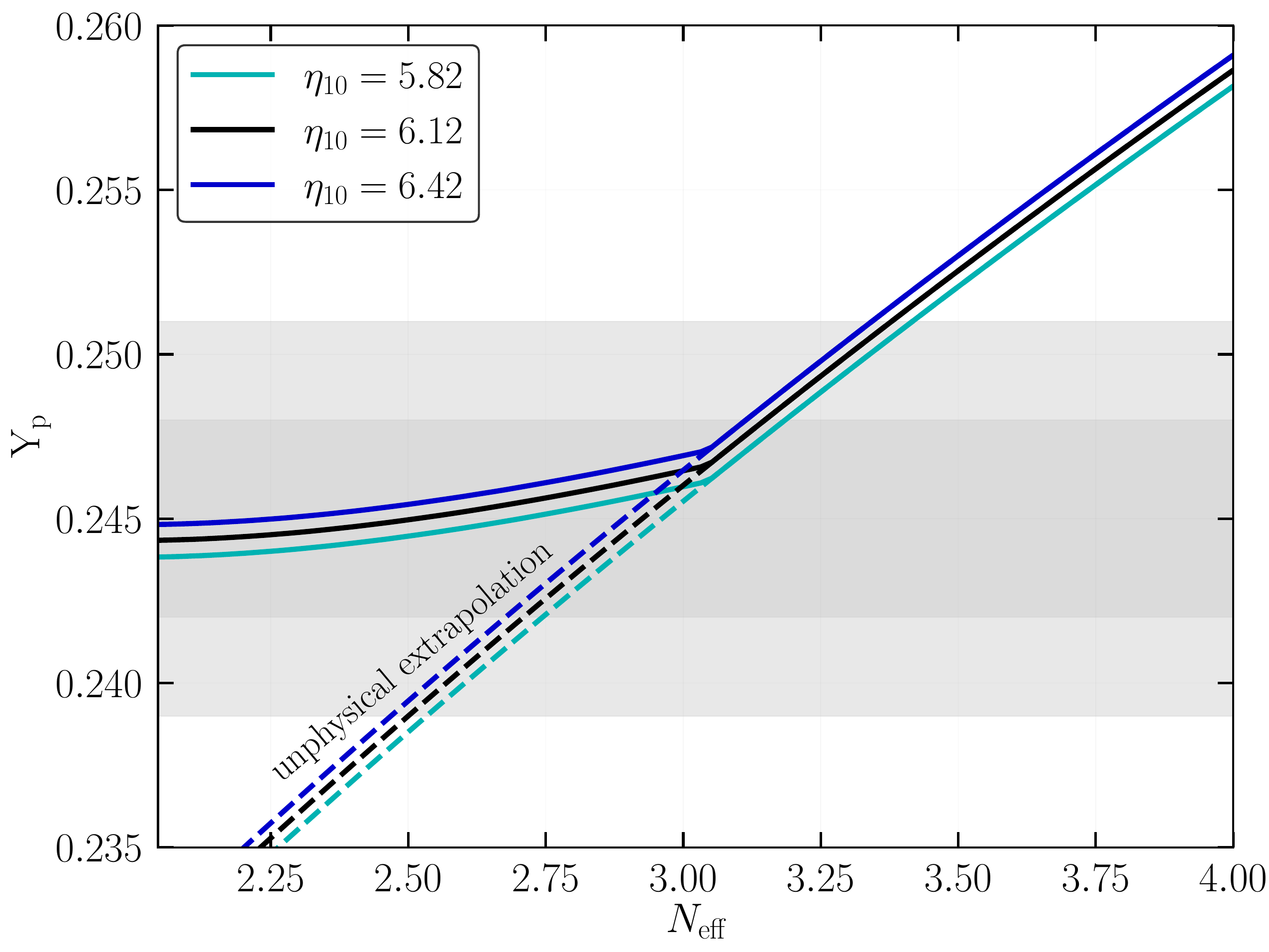}
    \includegraphics[width=0.48\textwidth]{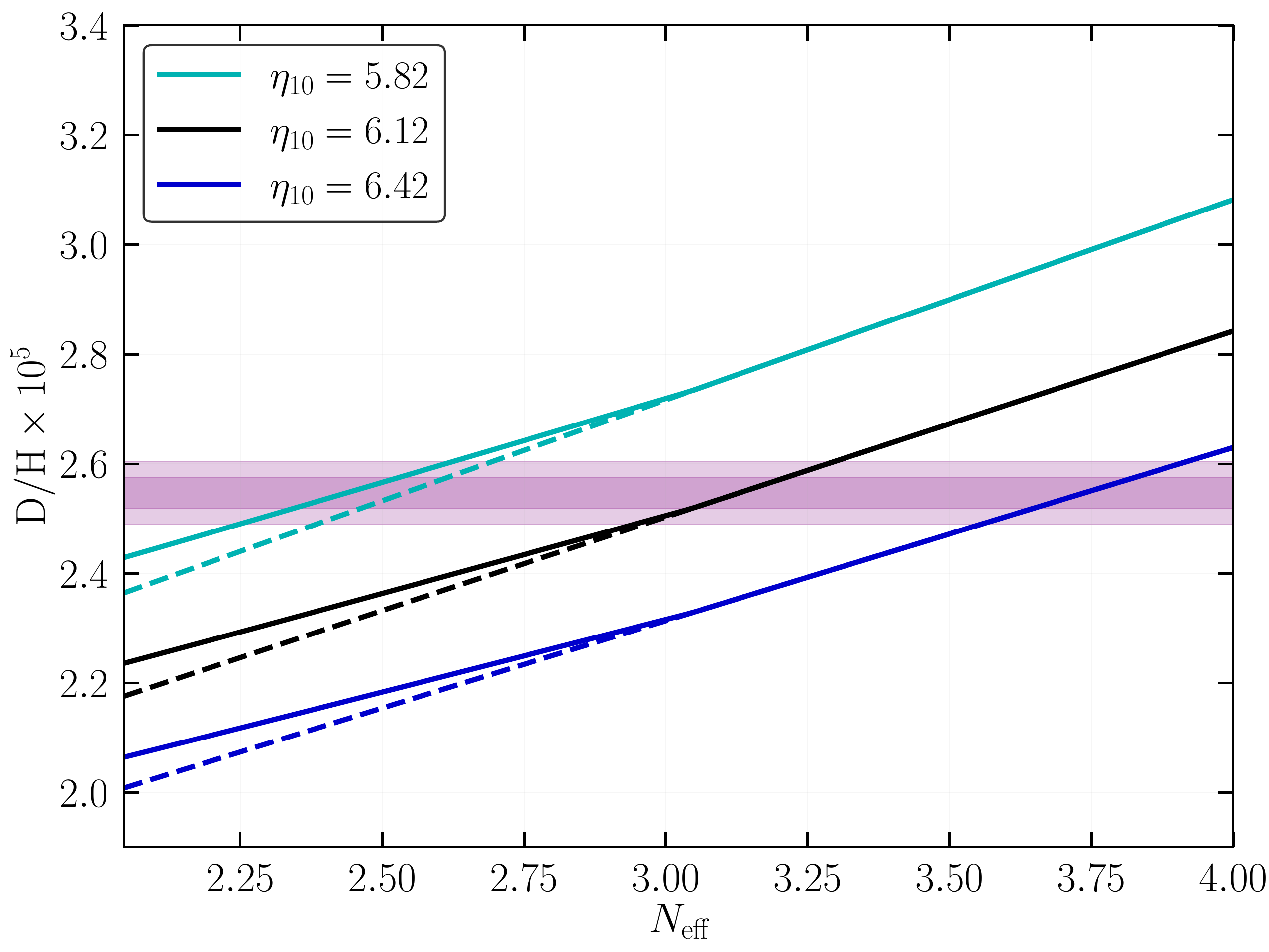}
    \caption{ 
    Primordial $\Yp$ (left) and $\D/\H$ (right) are depicted by solid lines as functions of $N_{\rm eff}$, for various $\eta_{10} = 5.82$ (cyan), $6.12$ (black), and $6.42$ (blue).
    Observational values of $\Yp$ and $\D/\H$ are taken from PDG, and their $1\sigma$ and $2\sigma$ uncertainties are depicted by the gray and purple bands for $\Yp$ and $\D/\H$, respectively.
    Dashed lines correspond to the conventional predictions obtained by the unphysical extrapolation of using Eq.\,\eqref{Eq:Tnu_ratio_DR}.
    }
    \label{Fig:Yp_and_D}
\end{figure*}

\section{Results}
\label{sec:results}
Fig.\,\ref{Fig:Yp_and_D} shows the details of how the primordial $^4\He$ mass fraction $\Yp\equiv \rho(^4\He)/\rho_b$ (left) and the deuterium-to-hydrogen ratio $\D/\H$ (right) are affected by the change of $N_{\rm eff}$.
Our results on $\Yp$ agree well with our linear relations \eqref{Eq:Xnfo_EI} and \eqref{Eq:Xnfo_DR} (see more details in Appendix.~\ref{App:Refined}).
Observational values are taken from the recommended values in particle data group (PDG), and their $1\sigma$ and $2\sigma$ uncertainties are depicted by the gray and purple bands for $\Yp$ and $\D/\H$, respectively.
Solid lines are the theoretical predictions, where we take $\eta_{10} \equiv \eta_b \times 10^{10} = 5.82$ (cyan), $6.12$ (black), and $6.42$ (blue), while the black line represents the best-fit value of $\eta_{10}$ from the Planck fitting\,\cite{Planck:2018vyg}.

At $\Delta N_{\rm eff} <0$, the solid lines are estimated by using Eq.~\eqref{Eq:Tnu_ratio}, while we also depict the conventional predictions by dashed lines, which are unphysical extrapolations of Eq.~\eqref{Eq:Tnu_ratio_DR}.
As shown in the figure, $\Yp$ loses its sensitivity on $N_{\rm eff}$ significantly at $\Delta N_{\rm eff}<0$.
At the same time, $\Yp$ is also insensitive to $\eta_{10}$, and therefore the measurement of $\Yp$ does not provide useful constraint on $N_{\rm eff}$ and $\eta_{10}$ when $\Delta N_{\rm eff}$ is negative.
Theoretical prediction of $\Yp$ in large ranges of $N_{\rm eff}$ and $\eta_{10}$ is still inside $1\sigma$ band.
Consequently, in the space of $\eta_{10}$ and $\Delta N_{\rm eff}<0$, we have only one effective measurement, $\D/\H$, and cannot specify preferred values of two fitting parameters.
This implies that the BBN analysis alone cannot rule out some parameter set at $\Delta N_{\rm eff}<0$ that would have been excluded by the conventional fitting; see, for instance, $\Delta N_{\rm eff} \simeq -0.5$ and $\eta_{10} = 5.82$ on the cyan line.

The left panel of Fig.\,\ref{Fig:Yp_and_D} also shows that it is difficult to decrease $\Yp$ by a negative $\Delta N_{\rm eff}$.
For instance, the EMPRESS collaboration\,\cite{Matsumoto:2022tlr} reported $\Yp^{\rm EMPRESS} = 0.2370^{+0.0034}_{-0.0033}$, which is about $2\sigma$ smaller than the PDG-recommended value~\eqref{Eq:Yp_obs}, and found that the standard BBN scenario is disfavored compared to $\Delta N_{\rm eff} \simeq -0.5$ by more than $2\sigma$, following the conventional treatment, i.e. using Eq.\,\eqref{Eq:Tnu_ratio_DR}.
However, as we have repeatedly argued, there is no realistic particle physics model that can take Eq.\,\eqref{Eq:Tnu_ratio_DR} with a negative $\Delta N_{\rm eff}$.
It is explicitly shown, in the left panel of Fig.\,\ref{Fig:Yp_and_D}, that such a small $\Yp$ cannot be realized in the electromagnetic injection scenario.
Note that hadronic injection secnario still tends to increase $\Yp$\ \cite{Reno:1987qw, Kohri:1999ex, Kohri:2001jx, Pospelov:2010cw}.
This problem may be resolved by a large lepton asymmetry with $N_{\rm eff} \simeq 3$ as shown in Refs.\,\cite{Matsumoto:2022tlr, Burns:2022hkq, Escudero:2022okz}; see also Refs.\,\cite{Kawasaki:2022hvx, Borah:2022uos, ChoeJo:2023cnx} as examples of realizing such a scenario.

We conduct the simple $\chi^2$ fitting, where we define and minimize
\bal
\chi^2 = 
\!\!\!\!\!\!
\sum_{X=\Yp, (\D/\H)} \!\!\!\!\!\!
\frac{
[X^{\rm (obs)}-
X^{\rm (mod)}(\eta_B, N_{\rm eff})]^2}
{\sigma_{X,{\rm obs}}^2+\sigma_{X,{\rm mod}}^2}.
\eal
The uncertainties of $\sigma_{X,{\rm obs}}$ and $\sigma_{X,{\rm mod}}$ come from the measurement and theoretical calculation, respectively.
Following Ref.~\cite{Matsumoto:2022tlr}, we take $\sigma_{\Yp,{\rm mod}}^2=(0.00003)^2 + (0.00012)^2$ and $\sigma_{(\D/\H),{\rm mod}}^2=(0.06)^2\times 10^{-10}$.
Our result can be summarized as in Fig.~\ref{fig:summary}.
We put an upper bound of $\Delta N_{\rm eff}$ at 95\,\% C.L. as
\bal
\Delta N_{\rm eff} < 0.74
\,.
\eal
We could not obtain its lower bound because a too small $N_{\rm eff}$ cannot be realized even by the electromagnetic injection, so we truncate our analysis at $N_{\rm eff}=2$.

\section{Conclusion}
\label{sec:discussions}

In this paper, we have pointed out the difficulty of implementing a consistent treatment of $N_{\rm eff}$ in BBN analysis.
The fundamental reason for this difficulty comes from the fact that neutrinos matter in BBN, while a negative $\Delta N_{\rm eff}$ necessarily modifies the neutrino sector in a model-dependent way.
The fitting results obtained in the conventional method (where the neutrino temperature is unmodified even at $\Delta N_{\rm eff} <0$) cannot be used for any particle physics model.

Accepting some model dependence, we have also introduced a scheme to conduct $N_{\rm eff}$ fitting for BBN, where we assume the dark radiation for $\Delta N_{\rm eff} >0$ and early electromagnetic injection for $\Delta N_{\rm eff}<0$.
This allows, at least, a physically sensible interpretation of the BBN fitting result.
However, as shown analytically in Sec.\,\ref{sec:analytic} and also depicted explicitly in Figs.\,\ref{fig:summary} and \ref{Fig:Yp_and_D}, $\Yp$ loses its sensitivity on $N_{\rm eff}$ at $\Delta N_{\rm eff}<0$, making the BBN fitting itself less useful for constraining $N_{\rm eff}$.

All these aspects indicate that $N_{\rm eff}$ is not a good effective fitting parameter in the BBN analysis. 

\vspace{0.2cm}
\noindent
{\bf Acknowledgement}
We thank Kazunori Kohri for helpful comments and discussion.
This work was supported by IBS under the project code, IBS-R018-D1.

\begin{appendix}

\section{A more refined analysis}
\label{App:Refined}

In this appendix, we present a more refined setup to justify the assumptions made in our analysis.
We consider a long-lived particle $\chi$ decays electromagnetically (photons or $e^\pm$) with its lifetime $\tau_\chi=\Gamma_\chi^{-1}$, mass $m_\chi$, and initial yield $Y_\chi$.
One may consider $\chi$ as an axion-like particle with $\chi F \tilde F$ interaction, or a CP-even scalar particle that has an effective operator $\chi F F$.
However, note that model details are not important here; we simply assume that the decay products instantly get thermalized into the photon sector. 
The goal of this appendix is to find conditions in terms of $\tau_\chi$ and $m_\chi$ under which our effective treatment of the negative $\Delta N_{\rm eff}$ is valid.

Since precise calculation is not our point, we built our own code with some simplifications.
Firstly, we take the evolution equations for the photon and electron sector energy density $\rho_{e\gamma}=\rho_\gamma+\rho_{e^\pm}$ and the neutrino sector energy densities $\rho_{\nu_\alpha}$ ($\alpha=e,\mu,\tau$) as
\bal
&\dot \rho_{e\gamma}+ 3H(\rho_{e\gamma}+p_{e\gamma})= \Gamma_\chi m_\chi n_\chi
-\sum_\alpha \Big( \frac{\delta \rho_{\nu_\alpha}}{\delta t} \Big),
\\
&\dot \rho_{\nu_\alpha}+4H\rho_{\nu_\alpha} = 
\Big( \frac{\delta \rho_{\nu_\alpha}}{\delta t} \Big).
\eal
where $\Big( \! \frac{\delta \rho_{\nu_\alpha}}{\delta t} \! \Big)$ are the energy transfer rates among neutrinos and photon sectors.
We approximate $\Big( \! \frac{\delta \rho_{\nu_\alpha}}{\delta t} \! \Big)$ by the analytic forms in Ref.\,\cite{Escudero:2018mvt}, where Maxwell-Boltzmann statistics was used for thermal average, while neutrino's temperature is obtained by $T_{\nu_\alpha} = ((\frac{\pi^2}{30}\frac{7}{4})^{-1} \rho_{\nu_\alpha})^{1/4}$.
Our code leads to $N_{\rm eff} \simeq 3.041$ in SBBN, which is only about $1$\,--\,$2\%$ deviation from more precise estimation $3.043$ -- $3.046$\,\cite{Mangano:2001iu, Mangano:2005cc, deSalas:2016ztq, EscuderoAbenza:2020cmq, Akita:2020szl, Cielo:2023bqp}.
For $n_\chi$, we solve the evolution equation,
\bal
\dot n_\chi + 3H n_\chi = -\Gamma_\chi n_\chi,
\eal
whose initial condition is set by the initial yield $Y_\chi$ at $T_i=30\,\MeV$.
We take $Y_\chi$ as a free parameter to scan $\Delta N_{\rm eff}$.
When $Y_\chi\neq0$, we should begin with a larger baryon density to compensate for the dilution coming from $\chi$ decay (the baryon-to-photon ratio $\eta_B$ is fixed at the current Universe).
For this, we simultaneously solve the evolution equation for $n_B$: $\dot n_B +3H n_B =0$.

As mentioned in the main text, the distortion of the neutrino distributions can be significant when a long period of $\chi$ domination appears.
Therefore, $m_\chi$ has to be smaller than about $10\,\GeV$.

The Hubble rate is obtained by $H^2=(8\pi/3m_{\rm pl}^2)(\rho_{e\gamma}+\sum_\alpha \rho_{\nu_\alpha}+m_\chi n_\chi)$, ignoring the baryon energy density, which is suppressed by the small $\eta_B m_N/T$.
Baryon energy density can be important when $T\lsim \eV$, far after the primordial abundances of light elements are frozen.

We take nuclear reaction rates from Refs.~\cite{Pisanti:2007hk, Consiglio:2017pot, Gariazzo:2021iiu} up to $^7{\rm Be}$.
This is sufficient to estimate the deuterium and $^4{\rm He}$ abundances since the abundances of heavier elements are negligible.

Numerically, our simplified code gives $Y_p=0.2453$ and $\D/\H=2.592\times 10^{-5}$ for SBBN with the neutron lifetime $\tau_n=878.4$ and the current baryon-to-photon ratio $\eta_B=6.12\times 10^{-10}$. 
Our results have deviations from other public codes up to $0.4\,\%$ for $Y_p$ and $4\,\%$ for $\D/\H$.
Although the latter deviation is actually larger than the observational uncertainty, failing to explain the observation, it is still approximately in the same order as the uncertainty coming from nuclear reaction rates\,\cite{Yeh:2020mgl}.

\begin{figure*}[t]
    \centering
    \includegraphics[height=0.21\textheight]{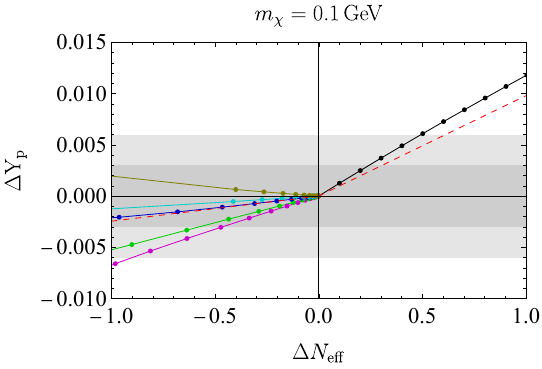}
    \includegraphics[height=0.21\textheight]{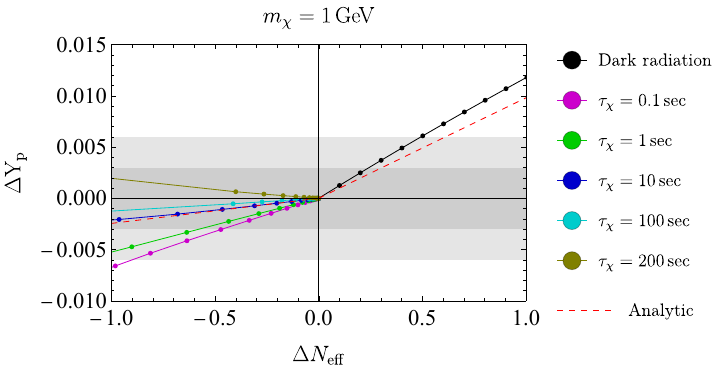}
    \caption{$\Delta N_{\rm eff}$ dependence of $\Delta Y_p$ for different $m_\chi=0.1\,\GeV$ (left) and $1\,\GeV$ (right).
    We take $\tau_\chi = 0.1\,\sec$ (purple), $1\,\sec$ (green), $10\,\sec$ (blue), $100\,\sec$ (cyan) and $200\,\sec$ (brown).
    The black dots and line represent the dark radiation case, and our analytic estimations in Eqs.~\eqref{Eq:Xnfo_EI} and \eqref{Eq:Xnfo_DR} are depicted by the dashed red line.
    Gray shaded regions correspond to the $1\,\sigma$ and $2\,\sigma$ uncertainties of observed $Y_p$.
    }
    \label{fig:Delta_Yp_appendix}
\end{figure*}

\begin{figure*}[t]
    \centering
    \includegraphics[height=0.21\textheight]{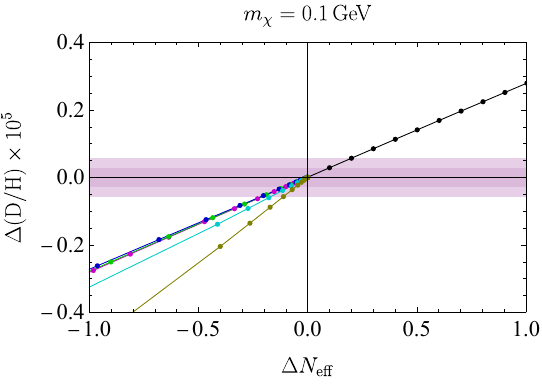}
    \includegraphics[height=0.21\textheight]{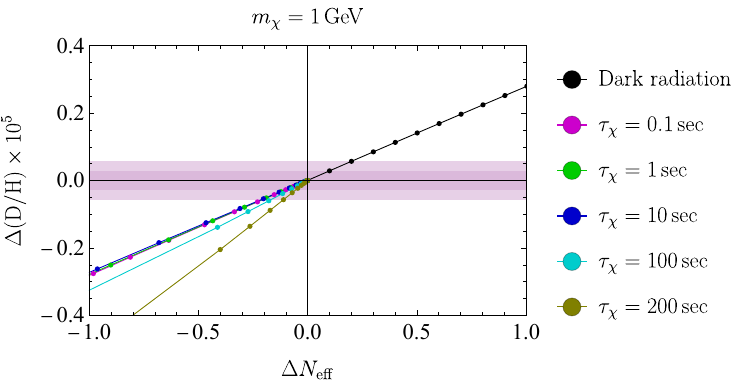}
    \caption{$\Delta N_{\rm eff}$ dependence of $\Delta (\D/\H)$ for different $m_\chi=0.1\,\GeV$ (left) and $1\,\GeV$ (right).
    We take $\tau_\chi = 0.1\,\sec$ (purple), $1\,\sec$ (green), $10\,\sec$ (blue), $100\,\sec$ (cyan) and $200\,\sec$ (brown).
    The black dots and line represent the dark radiation case.
    Purple shaded regions correspond to the $1\,\sigma$ and $2\,\sigma$ uncertainties of observed $(\D/\H)$.
    }
    \label{fig:Delta_DoverH_appendix}
\end{figure*}

Our results of $\Delta Y_p$ and $\Delta (\D/\H)$ for different $\Delta N_{\rm eff}$ are shown in Fig.~\ref{fig:Delta_Yp_appendix} and Fig.~\ref{fig:Delta_DoverH_appendix}, respectively.
Dots are our numerical evaluations, and lines are the interpolation of them.
In our scheme, $\Delta N_{\rm eff}$ is controlled by scanning $Y_\chi$, and $m_\chi$ is taken to be $0.1\,\GeV$ (left) and $1\,\GeV$ (right).
Different colors represent different $\tau_\chi = 0.1\,\sec$ (purple), $1\,\sec$ (green), $10\,\sec$ (blue), $100\,\sec$ 
(cyan), and $200\,\rm sec$ (brown).
The dark radiation case is depicted by the black dots, and our estimations in Eqs.~\eqref{Eq:Xnfo_EI} and \eqref{Eq:Xnfo_DR} are also shown by the dashed red line.
These results show good agreement with the analytic estimates and also Fig.~\ref{Fig:Yp_and_D} for $0.1\,\sec \lesssim \tau_\chi \lesssim 100\,\sec$.

When $\tau_\chi \gtrsim 100\,\sec$, the deuterium abundance starts deviating from others (see Fig.\,\ref{fig:Delta_DoverH_appendix}).
This is because the entropy injection occurs after the deuterium bottleneck temperature ($t_\D \sim 200\,\sec$), leading to a larger ``prior" baryon density with which nucleosynthesis proceeds (recall that the initial baryon yield is greater than the baryon yield in the current Universe after $\chi$ injects entropy, and also that $\D/\H$ is very sensitive to the baryon density).
Thus, it provides an upper limit of $\tau_\chi$ for the validity of our method.

The deviation in $Y_p$ at $\tau_\chi \lesssim 0.1\,\sec$ (see Fig.\,\ref{fig:Delta_Yp_appendix}) comes from different decoupling temperature of $\nu_e$ and $\nu_\mu$/$\nu_\tau$; the electron neutrino decouples after muon/tau neutrinos decouple because of the absence of muon/tau lepton in the background plasma around $T\sim \MeV$.
Thus, the dilution of muon/tau neutrinos due to $\chi$ decay starts earlier than that of electron neutrinos.
Consequently, the electron neutrino temperature is, in general, less affected by the dilution than the muon/tau neutrino temperature.

\begin{figure}[t]
    \centering
    \includegraphics[width=0.45\textwidth]{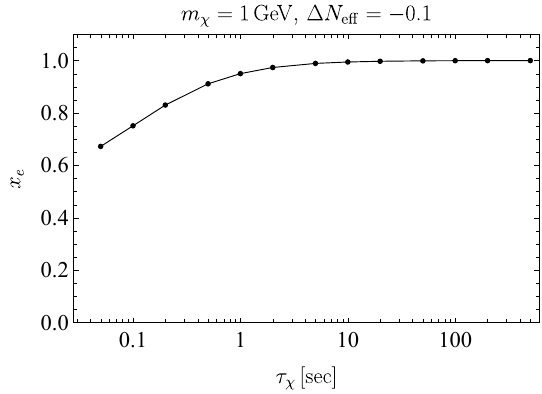}
    \includegraphics[width=0.45\textwidth]{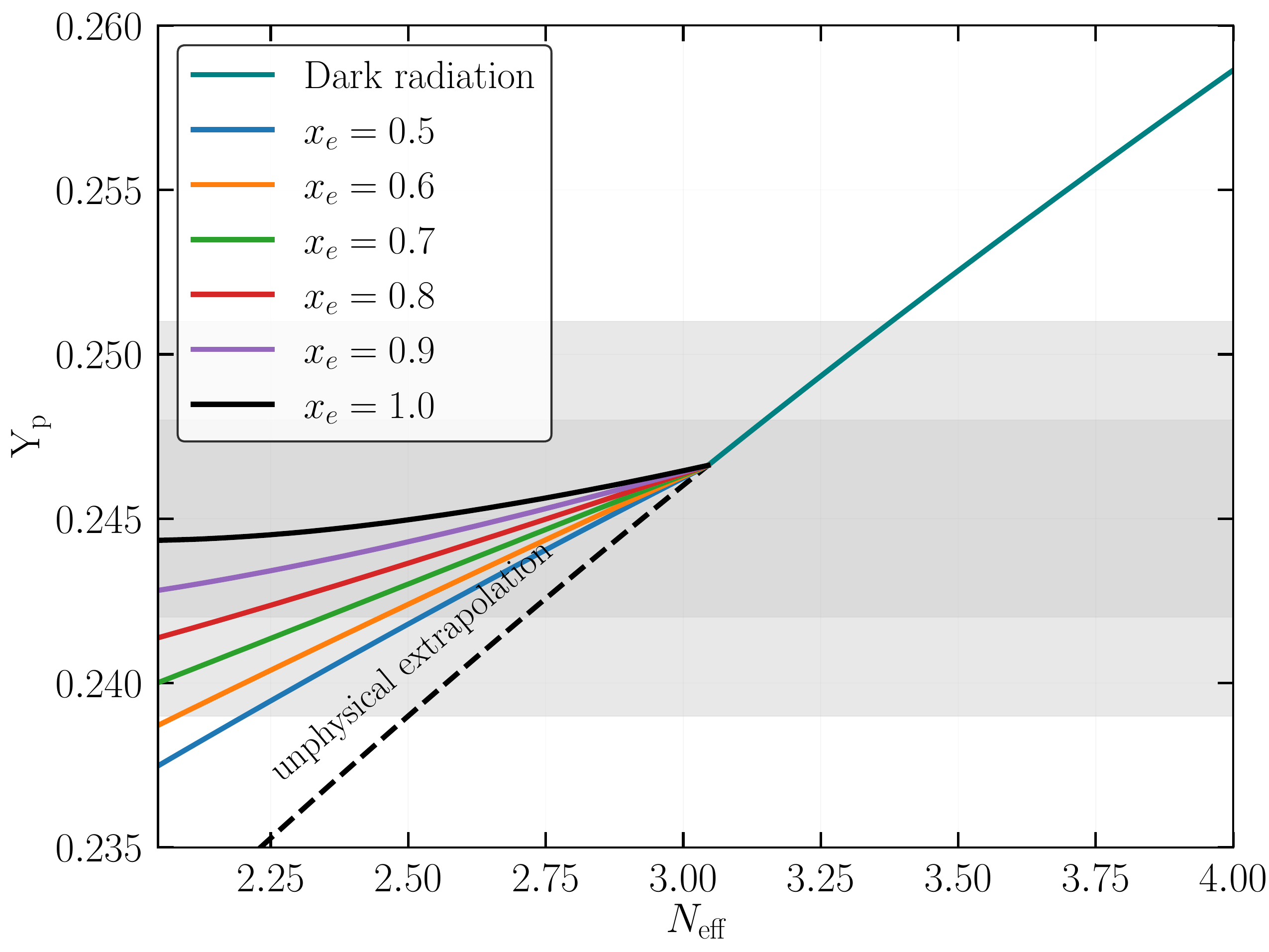}
    \caption{Top: $x_e$ values as a function of $\tau_\chi$ for $m_\chi=1\,\GeV$ and $\Delta N_{\rm eff}=-0.1$. 
    Bottom: Variation of $\Yp$ as a function of $N_{\rm eff}$ for different
    values of $x_e$. The dashed line denotes the unphysical extrapolation
    of dark radiation scenario for $\Delta N_{\rm eff} < 0$.
    }
    \label{fig:xe_appendix}
\end{figure}

This can be quantitatively understood in terms of the $x_e$ parameter introduced in Eq.\,\eqref{Eq:Tnu_ratio}.
$x_e$ can be rewritten as
\bal
\frac{x_e}{N_{\rm eff}^\SBBN} =
\frac{1}{\Delta N_{\rm eff}} \Big( \frac{\rho_{\nu_e}}{\rho_{\nu_e}^\SBBN}-1 \Big)
\simeq \frac{\rho_{\nu_e}-\rho_{\nu_e}^\SBBN}{\rho_{\nu_e}-\rho_{\nu_e}^\SBBN+2(\rho_{\nu_\mu}-\rho_{\nu_\mu}^\SBBN)}
,
\eal
meaning that $x_e/3$ represents the fraction of $\Delta N_{\rm eff}$ contributed by $\nu_e$.
If $x_e=1$, all flavors contribute equally, while $x_e<1$ means the $\nu_e$'s contribution is small.
In the top panel of Fig.~\ref{fig:xe_appendix}, we show its numerical values as a function of $\tau_\chi$ for $m_\chi=1\,\GeV$ and $\Delta N_{\rm eff}=-0.1$.
As one can see, when $x_e$ starts deviating from one at $\tau_\chi\lsim O(0.1)\,\sec$.
To apply our effective method introduced in this case, one needs to adjust $x_e$, and the bottom panel of Fig.\,\ref{fig:xe_appendix} shows our result on it.
Note, however, as $\tau_\chi$ decreases, $|\Delta N_{\rm eff}|$ gets exponentially suppressed $\propto \exp(-t_{\nud}/\tau_\chi)$, and BBN loses its sensitivity.

As a final remark of this appendix, we point out that the physical situation is very different for $\tau_\chi \gtrsim 10\,\sec$ from what is described in the main text.
When $\chi$ decays with $\tau_\chi \gtrsim 10\,\sec$, the neutron should already be frozen, and the only contribution from the presence of $\chi$ would be a slight modification of the Hubble rate.
This modification is proportional to a given $\Delta N_{\rm eff}$, but it is suppressed at $t<\tau_\chi$ as the energy density of $\chi$ is matter-like.
This is the actual reason why $Y_p$ is insensitive to $\Delta N_{\rm eff}$ at a large $\tau_\chi$.

Nevertheless, our numerical study presented in this appendix shows that our ``effective" treatment can cover a wide range of $0.1\,\sec \lesssim \tau_\chi \lesssim 100\,\sec$ and $0.1\,\GeV \lesssim m_\chi \lesssim 10\,\GeV$.
We expect that our method can still be valid, but the validation will require a more careful treatment of neutrino distributions.

\end{appendix}


%

\end{document}